\documentclass{ifacconf}

\usepackage{natbib}        
\usepackage{siunitx}
\usepackage{mathrsfs}
\usepackage{color}
\usepackage{enumerate,g	raphicx,epstopdf}
\usepackage{amsmath,amssymb,bm}
\usepackage{cite}
\usepackage{mathtools}
\usepackage{array}
\usepackage{algpseudocode,algorithm,algorithmicx}
\usepackage{booktabs}
\usepackage[hyphens]{url}
\usepackage{lipsum}
\usepackage[deletedmarkup=sout,authormarkup=superscript]{changes}
\usepackage[normalem]{ulem}
\usepackage{flushend}

\newtheorem{rmk}{Remark}

\newtheorem{lmm}{Lemma}

\newenvironment{proof}[1][Proof:]{\begin{trivlist}
\item[\hskip \labelsep {\itshape #1}]}{\end{trivlist}}

\newcommand{\R}{\mathbb{R}}

\newcommand{\mc}[1]{\mathcal{#1}}
\newcommand{\bb}[1]{\mathbb{#1}}

\newcommand{\col}{\mathrm{col}}

\definechangesauthor[name={Giuseppe}, color=blue]{GB}
\definechangesauthor[name={Varsha}, color=red]{VB}

\begin{document}

\begin{frontmatter}

\title{Designing Fairness in Autonomous Peer-to-peer Energy Trading}
\thanks[footnoteinfo]{This research is supported by the SNSF through NCCR Automation (Grant Number 180545). Email addresses: \{varsha.behrunani,philipp.heer\}@empa.ch, airvine@student.ethz.ch, \{bvarsha, gbelgioioso, jlygeros, doerfler\}@ethz.ch.}

\author[First,Second]{Varsha N. Behrunani} 
\author[Second]{Andrew Irvine}
\author[Second]{Giuseppe Belgioioso}
\author[First]{Philipp Heer} 
\author[Second]{John Lygeros}
\author[Second]{Florian D\"{o}rfler}

\address[First]{Empa, Urban Energy Systems Laboratory, D\"{u}bendorf, Switzerland}
\address[Second]{Automatic Control Laboratory, ETH Z\"{u}rich, Switzerland}

\begin{abstract}
Several autonomous energy management and peer-to-peer trading mechanisms for future energy markets have been recently proposed based on optimization and game theory. In this paper, we study the impact of trading prices on the outcome of these market designs for energy-hub networks. We prove that, for a generic choice of trading prices, autonomous peer-to-peer trading is always network-wide beneficial but not necessarily individually beneficial for each hub. Therefore, we leverage hierarchical game theory to formalize the problem of designing locally-beneficial and network-wide fair peer-to-peer trading prices. Then, we propose a scalable and privacy-preserving price-mediation algorithm that provably converges to a profile of such prices. Numerical simulations on a 3-hub network show that the proposed algorithm can indeed incentivize active participation of energy hubs in autonomous peer-to-peer trading schemes. 
\end{abstract}


\end{frontmatter}


\section{Introduction}

 
Technology improvements for multi-generation and storage systems coupled with an increased penetration of renewable energy sources has led to an unprecedented rise in distributed energy resources, in the form of multi-energy hubs and prosumers, namely, energy consumers with storage and production capabilities. Energy hubs integrate multiple energy carriers, production, conversion, and storage units. Prosumers and energy hubs can supplement traditional utilities to supply demand and play a vital role for the energy balance of the grid while also enriching energy efficiency, and maximizing utility of renewable resources, therefore, decreasing overall energy costs and carbon footprint~\citep{Geidl:2007a}. However, this shift also requires markets to evolve from a hierarchical and centralized to a decentralized design that can enable active participation of prosumers.

Peer-to-peer (P2P) trading is an emerging feature of new distributed market designs that allows prosumers to directly share energy, that can be used to match demands locally or to reduce the reliance on the grid as well as the overall energy cost. In addition to the positive economic and environmental impact, P2P trading has also the potential to reduce peak demand and reserve requirements, thus lowering the need for expanding energy infrastructures and energy imports~\citep{Wayes:2020}. 
 
Several coordination mechanisms have been recently proposed in the literature to facilitate energy management and P2P trading in energy communities. Existing works formulate the problem either via multi-agent optimization \citep{Sorin:2019,baroche2019prosumer,moret2020loss} or noncooperative games \citep{le2020peer,Cui:2020,belgioioso2022operationally}.
In \citep{Sorin:2019,baroche2019prosumer, moret2020loss}, autonomous P2P trading is cast as a collective optimization problem and Lagrange relaxation-based methods are used to find the optimal operational set points in a distributed manner. In \citep{le2020peer, Cui:2020} each prosumer has local decoupled objectives and the energy trading is incorporated as coupling reciprocity constraints, resulting in a game with coupling constraints. In addition to bilateral trading, \citep{belgioioso2022operationally} also considers trading with the main grid, and includes network operational constraints (e.g., power flows, line capacities) and system operators in the model.
 
Bilateral trading prices play a key role in defining the outcome of all the aforementioned market models. In fact, naive choices of bilateral trading prices may induce undesired congestions in the power grid or disproportionate costs for the prosumers \citep{le2020peer}. To incentivize active participation in the market model, it is therefore crucial to ensure that individual market participants reap the network-wide benefits of autonomous trading.
 
The development of effective pricing schemes has been already considered in different works. In \citep{le2020peer}, the effect of P2P trading prices is investigated by studying energy preferences and marginal prices (dual variables) connected to the coupling trading reciprocity constraints. In \citep{Fan:2018}, a bargaining game is designed to distribute the benefits of P2P trading equally to all hubs. In \citep{Wang:2020}, the transactive prices of each hub are derived based on their internal operation strategy to ensure that their costs are recovered. In \citep{DARYAN:2022}, a two-stage mechanism is proposed to obtain the optimal payment that incentivizes active participation in the market design. 
Finally, \citep{Sorin:2019} considers product differentiation and preferences to define effective transaction prices. 

In this paper, we leverage hierarchical game theory and distributed optimization
to design a novel scalable, locally-benefical, and fair pricing scheme for autonomous peer-to-peer trading in energy hub networks that incentivizes active participation. Our contribution is three-fold:

\begin{enumerate}[(i)]
\item We formulate the problem of designing locally beneficial and network-wide fair trading prices as a bilevel game. At the lower level, the ``optimal" energy trades and operational setpoints for the energy hubs are formulated as interdependent economic dispatch problems. 
At the upper level, the desired P2P trading prices are defined as the minimizers of a fairness metric (namely, the sample variance of the local cost reductions of the hubs) which depends on the optimal setpoints of the lower-level dispatch problem.

\item We leverage the special structure of the bilevel game to decouple the two levels and design an efficient 2-step solution algorithm. In the first step, an ADMM-based algorithm is used for distributively solving the economic dispatch problems. In the second step, a semi-decentralized price mediation algorithm is used to compute fair P2P trading prices in a scalable way. 

\item We illustrate and validate the proposed autonomous P2P trading and pricing mechanism via extensive numerical simulations on a 3-hub network, using realistic models of energy hubs and demand data.

\end{enumerate}

%

\section{Modelling the hubs}
\begin{figure}[t]
\begin{center}
\includegraphics[width=8cm]{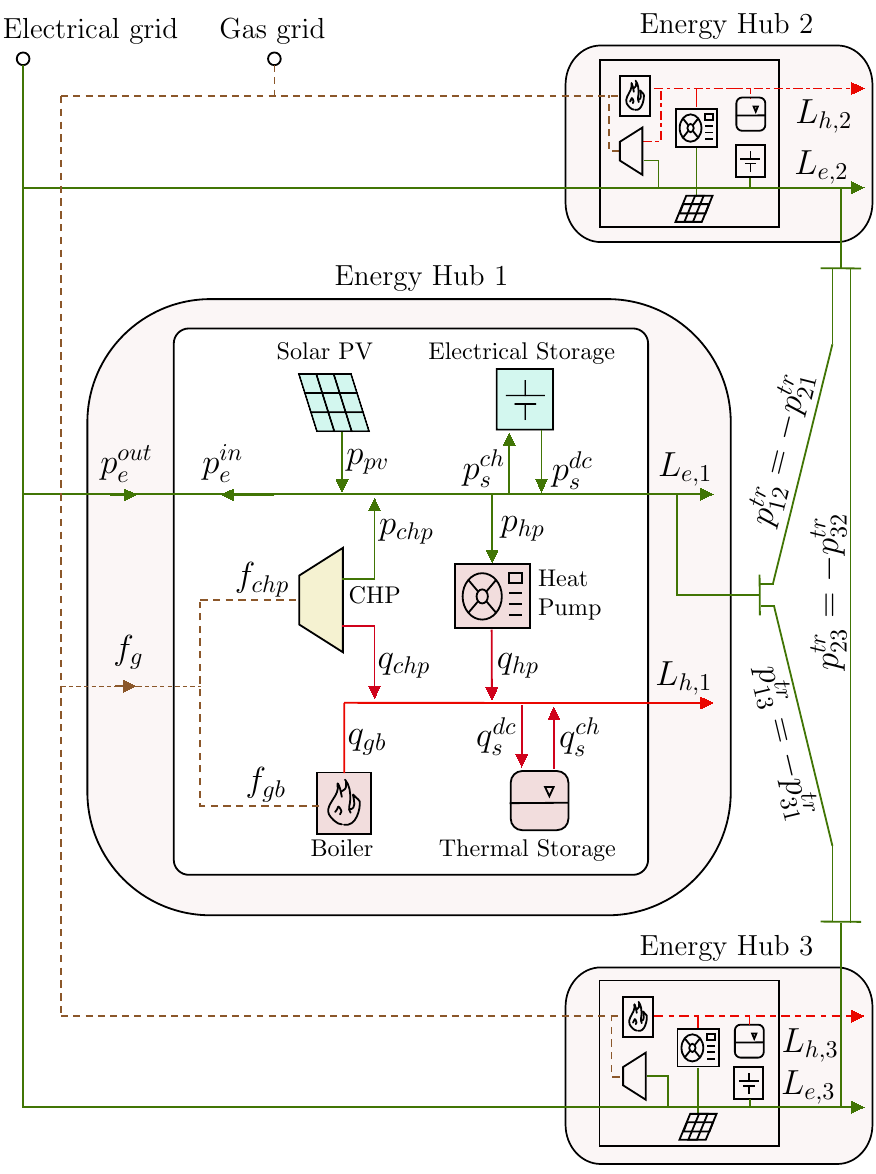}
\caption{A network of three interconnected energy hubs. Each hub can import energy from the electricity (green) and gas (brown) grids, and can feed-in electricity to the electricity grid; additionally, each hub can also trade electrical energy with the other hubs.
} 
\label{fig:ehub_network}
\end{center}
\end{figure}

We consider a network of $N$ interconnected energy hubs, labeled by $i \in \mathcal{N}:=\{
1,\ldots,N\}$. Each hub is connected to the electricity and gas grid, and can trade electrical energy with the other hubs via the electricity grid. As an example, a system of three hubs in illustrated in Fig.~\ref{fig:ehub_network} and will be used to fix ideas throughout the paper.

To fulfil its electrical and thermal demand, each hub is equipped with different energy conversion and storage devices that draw electricity and gas directly from the grid. The heating devices can include gas boilers, heat pumps, as well as thermal energy storage. Electricity can be locally produced via photovoltaic (PV), Combined Heat and Power (CHP) and micro-CHP ($\mu$CHP) devices and locally stored using batteries. In addition to the heating and electricity demand, cooling demand may also be present which is not considered in this work. It can be added with suitable devices, such as chillers, ice storage, and HVAC, without conceptual changes to our model.

The electricity grid acts as an infinite source and sink, namely, electricity can be directly drawn from the grid and excess electrical energy produced in the hub can be fed to the grid. Additionally, the hubs can exchange energy via peer-to-peer (or bilateral) trading through the grid. The hubs are connected to the heating and electricity demand via a downstream distribution network. The demands of all the downstream entities supplied by each hub are aggregated into a single demand. For the sake of simplicity, in this study, we assume that a perfect forecast is available for this demand. Similarly, we assume that a perfect forecast for the temperature and the solar radiation are also available. Forecast uncertainties can play a crucial role in optimally operating energy hubs and integrating it into our model is a topic of current work.

In the next sections, we provide an overview of the models used for the components in the energy hubs.

\subsection{Energy Conversion}

\subsubsection{Combined Heat and Power (CHP):} 
\begin{figure}
\begin{center}
\includegraphics[width=8cm]{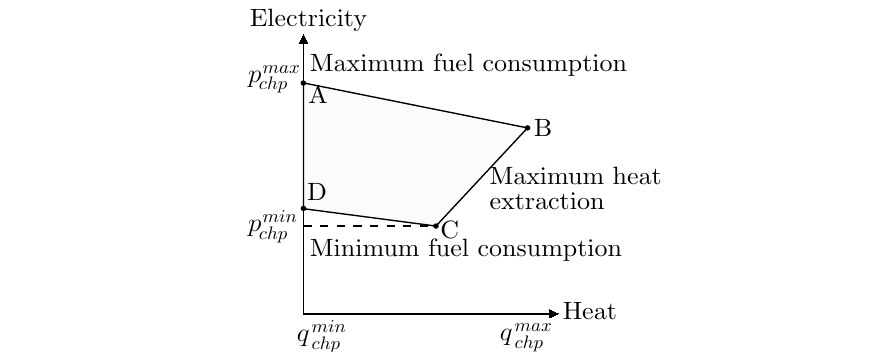}
\caption{Feasible region of combined heat and
power(CHP)} 
\label{fig:chp_model}
\end{center}
\end{figure}

The CHP simultaneously generates heat and power using natural gas.
The output is limited by its feasible operation region (Figure~\ref{fig:chp_model}) as defined by a polyhedron with vertices A-D and its corresponding electrical and thermal output, $p_{\text{a}}$, $p_{\text{b}}$, $p_{\text{c}}$, $p_{\text{d}}$ and $q_{\text{a}}$, $q_{\text{b}}$, $q_{\text{c}}$, $q_{\text{d}}$, respectively~\citep{navarro:2018, alipour:2014}. The electrical and thermal output of the CHP for hub $i$ are $p_{\text{chp,i}}$ and $q_{\text{chp,i}}$, respectively, characterized as a convex combination of the vertices with weights $w_{\text{a,i}}$, $w_{\text{b,i}}$, $w_{\text{c,i}}$, and $w_{\text{d,i}}$, respectively. The fuel consumed by the CHP unit, $f_{\text{chp}}$, depends only on the electrical output subject to the fuel efficiency, $\eta_{\text{chp}}$. CHP is modelled by the following equations:
\begin{equation}
\label{CHP_output}
\begin{aligned}
p_{\text{chp,i}} &= \sum_{j \in \mc K} w_\text{j,i} \cdot p_\text{j,i}\\
q_{\text{chp,i}} &= \sum_{j \in \mc K} w_\text{j,i} \cdot q_\text{j,i}\\
f_{\text{chp,i}} &= \frac{1}{\eta_{\text{chp,i}}} \cdot p_{\text{chp,i}}\\
1 &= \sum_{j \in \mc K} w_\text{j,i}\\
0 &\leq w_\text{j,i} \leq 1 \ \ \ \mc K=\{\text{A,B,C,D}\}
\end{aligned}
\end{equation}
\subsubsection{Heat Pump (HP):} The heat pump uses electricity to extract heat from the ground. The relation between heat pump electrical input $ p_{\text{hp,i}}$ and thermal output $q_{\text{hp,i}}$ defined by its coefficient of performance (COP) is given by
\begin{equation}
\label{GSHP}
\begin{aligned}
q_{\text{hp,i}} &= \text{COP} \cdot p_{\text{hp,i}}.
\end{aligned}
\end{equation}

\subsubsection{Gas boiler (GB):} The gas boiler uses natural gas, $f_{\text{gb,i}}$, to generate heat, $q_{\text{gb,i}}$. The thermal output of the boiler and its  efficiency, $\eta_{\text{gb,i}}$, is modelled by 
\begin{equation}
\label{CHP_fuel}
\begin{aligned}
q_{\text{gb,i}} &= \eta_{\text{gb,i}} \cdot f_{\text{gb,i}}.
\end{aligned}
\end{equation}

\subsubsection{Solar Photovoltaic (PV):} The energy output of the solar photovoltaic system at any time, $p_{\text{pv,i}}$, is quantified by the incident solar irradiance $I_{\text{solar}}$ $[\text{kW/m}^2]$ along with the total area $a_{\text{pv}}$ $[\text{m}^2]$ and the efficiency $\eta_{\text{pv}}$~\citep{Skoplaki:2009}. The solar irradiance $I_{\text{solar}}$ depends on the forecast of the global solar irradiation (which is assumed to be known) and the orientation of the PV on the building. 
\begin{equation}
\label{CHP_fuel}
\begin{aligned}
p_{\text{pv,i}} &= \eta_{\text{pv,i}} \cdot I_{\text{solar,i}} \cdot a_{\text{pv,i}}
\end{aligned}
\end{equation}

The output of all energy converters are also limited by the following capacity constraints:
\begin{equation}
\label{eq:cLimits}
\begin{aligned}
p^{\text{min}}_{\text{j,i}} &\leq p_{\text{j,i}}\leq p^{\text{max}}_{\text{j,i}} \ \ \ \text{j}=\{\text{pv, chp}\}.\\
q^{\text{min}}_{\text{k,i}} &\leq q_{\text{k,i}}\leq q^{\text{max}}_{\text{k,i}} \ \ \ \text{k}=\{\text{gb, hp, chp}\}.
\end{aligned}
\end{equation}

\subsection{Energy Storage System}
The dynamics of the electrical storage (ES) is modelled by the following discrete-time linear time-invariant system:
\begin{equation}
\label{storage}
p_{\text{s,i}}(h) = \gamma_{\text{e,i}} \cdot p_{\text{s,i}}(h-1) +  \eta_{\text{e,i}} \cdot p^{\text{ch}}_{\text{s,i}}(h)  - \left(\frac{1}{\eta_{\text{e,i}}}\right) \cdot p^{\text{dc}}_{\text{s,i}}(h),
\end{equation}
where $h\in \mathcal{Z}_{\geq 0}$ is the time index,  $p_{\text{s}}$ is the battery state of charge, $p^{\text{dc}}_{\text{s,n}}$ and  $p^{\text{ch}}_{\text{s}}$ are the energy discharged and charged into the battery, and $\gamma_{\text{e}}$ and $\eta_{\text{e}}$ are the standby and cycle efficiency.

The storage levels must satisfy the battery capacity limits 
\begin{equation}
\label{eq:stLimits}
p^{\text{min}}_{\text{j,i}} \leq p_{\text{j,i}}\leq p^{\text{max}}_{\text{j,i}} \ \ \ \text{j}=\{\text{s, dc, ch}\}.
\end{equation} 
The heat storage (TS) dynamics are modelled analogously with the corresponding state of charge, $q_{\text{s}}$, the heat discharged and charged into the thermal storage, $q^{\text{dc}}_{\text{s}}$ and  $q^{\text{ch}}_{\text{s}}$, and standby and cycle efficiency,  $\gamma_{\text{h}}$ and $\eta_{\text{h}}$, respectively. Storage units for thermal energy storage consists of devices such as borehole field, water tanks, etc. 


\subsection{Network modelling}
The network and internal connections describe the input-output equations of different energy carriers. 
For hub $i$, the energy balance constraint of electricity and heat read:
\begin{align}
\nonumber
L_{\text{e,i}} &=  p_{\text{chp,i}} + p_{\text{pv,i}} - p_{\text{hp,i}} + \left(p^{\text{out}}_{\text{e,i}} - p^{\text{in}}_{\text{e,i}}\right)   + \left( p^{\text{dc}}_{\text{s,i}}  - p^{\text{ch}}_{\text{s,i}}  \right)  \\ &+ \sum_{j\in \mc N \setminus\{i\}} p_{ij}^{\text{tr}}\ , \\
L_{\text{h,i}} &= q_{\text{chp,i}} + q_{\text{gb,i}} + q_{\text{hp,i}}  + \left( q^{\text{dc}}_{\text{s,i}}  - q^{\text{ch}}_{\text{s,i}}  \right),
\end{align}
where $L_{\text{e,i}}$ and $L_{\text{h,i}}$ are the electrical and thermal demand of the energy hub $i$ respectively, and $p^{\text{out}}_{\text{e,i}}$ and $p^{\text{in}}_{\text{e,i}}$ are the energy imported and fed into electricity grid respectively; While a district heating network may also be present in some places, it is not included here. In the absence of a heating grid, we assume demand can be met exactly at all times by conversion or storage.

The energy hubs can also trade electrical energy amongst each other. The power traded between hub $i$ and hub $j$ is $p^\text{tr}_{ij}$. The value is positive if the energy is imported by the hub $i$ and negative otherwise. The total energy exchanged between hub $i$ and the other hubs is $\sum_{j\in \mc N \setminus \{i\}}p_{ij}^{\text{tr}}$. 
Trading agreement is enforced via the so-called reciprocity constraints~\citep{Baroche:2019}\begin{subequations}
\begin{align}
\label{reciprocity}
p_{ij}^{\text{tr}} + p_{ji}^{\text{tr}} = 0, \ \ \  &\forall i,j\in \mc N, \ i \neq j,\\
\label{eq:TrLimits}
 p_{ij}^{\text{tr}} \leq \kappa_{ij}, \ \ \ &\forall i,j \in \mc N, \ i \neq j,
\end{align}
\end{subequations}
The reciprocity constraints \eqref{reciprocity} ensures that the energy exported from hub $i$ to hub $j$ is the same as the energy imported by hub $j$ from hub $i$; additionally, the constraints \eqref{eq:TrLimits} limit the trade between hubs where $\kappa_{ij}$ is the maximum that can be traded between the hubs $i$ and $j$. In this study, we assume that each hub can trade with all other hubs. Specific trading networks can be defined by restricting some of the trading limits \eqref{eq:TrLimits} to zero. 


\section{Autonomous P2P Trading}

\subsection{Economic Dispatch as a Noncooperative Game}
\label{subsec:EDasgame}
For each hub, the economic dispatch problem consists of choosing the local operational set points $p_i$, over a horizon $\mc H := \{1,\ldots,H \}$, to minimize its energy cost. The cost of each hub $i\in \mc N$ is the sum over all costs of its available assets (including the energy exchanged with the electricity grid), $\ell \in \mc{A}_i=\{\text{chp, gb, gshp, pv, grid}\}$ (namely, CHP unit, gas boiler, solar photovoltaic, electricity grid etc.), and the bilateral trades with the other hubs in the network. %

We model the cost of each asset $\ell \in \mathcal{A}_i$ as a strongly convex function $f_i^\ell(p_i^\ell)$, where $p_i^\ell \in \mathbb{R}^H$ is the vector of setpoints of asset $\ell$ over the horizon $\mc H$. Typical choices in the literature are quadratic and linear functions \citep{baroche2019prosumer, moret2020loss, le2020peer}. The total cost of bilateral trades with hub $j$ is given by 
\begin{align}
\label{eq:tradP}
c_{(i,j)}^\top p_{ij}^{\text{tr}} + \gamma {\| p_{ij}^{\text{tr}} \|}^2_2,
\end{align}
where $p_{ij}^{\text{tr}} \in \mathbb{R}^H$ collects the trades with hub $j$ over $\mc H$, $c_{(i,j)} \in \mathbb{R}^H$ defines the prices of the bilateral trades with hub $j$, while $\gamma$ is a marginal trading tariff imposed by the grid operator to use the network for bilateral trades. 



%
Overall, the economic dispatch problem of each hub $i$ can be compactly written as the following convex program:
\begin{subequations}
\label{eq:EDPi}
\begin{align}
\min_{p_i} \quad & 
 \overbrace{ \sum_{\ell \in \mc A_i} f^\ell_i\left(p_i^\ell \right) + \sum_{j \in \mc N} \left(
c_{(i,j)}^\top p^{\text{tr}}_{ij} +
\gamma {\|p^{\text{tr}}_{ij}\|}^2_2
\right)}^{=:J_i\left(p_i, c_i\right)}
\\
\text{s.t.} \quad
\label{eq:Constr1}
&  p_i \in \mathscr{P}_i \\
\label{eq:RC}
& p_{ij}^{\text{tr}} + p_{ji}^{\text{tr}} = 0, \ \ \  \forall j\in \mc N\setminus\{i \},
\end{align}
\end{subequations}
where $p_i$ collects all the decision variables of hub $i$ (local set points $p_i^\ell$, and import/exports from/to other hubs $p^{\text{tr}}_{ij}$), and $\mathscr{P}_i:=\{
p_i ~|~
\eqref{eq:stLimits}-\eqref{eq:TrLimits} \text{ hold}\}$ all its operational constraints; finally, the cost function $J_i(p_i,c_i)$ combines the costs of all the local assets and the bilateral trades, and its parametric dependency on the bilateral trading prices $c_i = (c_{(i,1)},\ldots,c_{(i,N)}) \in \mathbb R^{(N-1)H}$ is made explicit.

Note that the economic dispatch problems \eqref{eq:EDPi} are coupled via the reciprocity constraints \eqref{eq:RC} that enforce agreement on the bilateral trades. Therefore, the collection of $N$ parametric inter-dependent optimization problems \eqref{eq:EDPi} constitutes a game with coupling constraints \citep{facchinei2010generalized}.
A relevant solution concept for the game (\ref{eq:EDPi}) is that of generalized Nash equilibrium (GNE). Namely, a feasible action profile $p^* = (p_1^*,\ldots, p_N^*)$ for which no agent can reduce their cost by unilaterally deviating \citep[Def.~1]{belgioioso2022distributed}.
%
%
Here, we focus a special subclass of GNEs known as variational GNEs (v-GNEs) and characterized by the solution set, $S(c)$, of the variational inequality $\textrm{VI}(F(\cdot,c),\mathscr{P})$. Namely, the problem of finding a vector $p^* \in \mathscr{P}$ such that%
\begin{equation}
\label{eq:VIprob}
{F(p^*,c)}^{\top}(p-p^*) \geq 0, \quad \forall p \in \mathscr{P},
\end{equation}
where $\mathscr{P}:=\{p~|~\eqref{eq:Constr1}-\eqref{eq:RC} \text{ hold for all } i \in \mc N \}$, and $F$ is the so-called pseudo-gradient mapping, defined as
$$F(p,c)= \col(\nabla_{p_1} J_1(p_1,c_1),\ldots,\nabla_{p_N} J_N(p_N,c_N))$$
and parametrized by the bilateral trading prices $c$. 
This subclass of GNEs enjoys the property of ``economic fairness", namely, the shadow price (dual variable) due to the
presence of the coupling reciprocity constraints is the same for each hub \citep{facchinei2010generalized}. 

Interestingly, since the cost functions $J_i$ are decoupled, solutions of the variational GNEs of \eqref{eq:EDPi} correspond to the minimizers of the social cost problem
\begin{align}
\label{eq:SWP}
\left\{ \min_{p}  \; \sum_{i \in \mc N}
J_i\left(p_i, c_i\right) \quad \text{s.t.} \quad
 p \in \mathscr{P} \right\} =: W(c).
\end{align}
This equivalence directly follows by comparing the KKT conditions of $\textrm{VI}(F(\cdot,c),\mathscr{P})$ with that of the social cost problem \eqref{eq:SWP}, and was noted in a number of different works \citep{le2020peer, moret2020loss}. In the remainder, we exploit this connection in a number of ways.

First, we prove that the trading game \eqref{eq:EDPi} has a unique variational GNE which does not depend on the specific choice of the bilateral trading prices.
\begin{lmm}
\label{lmm:Sens}
Let $S(\cdot)$ be the price-to-variational GNE mapping, i.e.,
$S(c)=\{  p^* ~|~ {F( p^*,c)}^{\top}(p- p^*) \geq 0, \; \forall p \in \mathscr{P}
\}$. Then, $S(c) = \{ p^*\}$ for all price profiles $c \in \R^{N(N-1)H}$, for some unique profile $p^*\in \mathscr{P}$ independent of $c$.
\end{lmm}
\begin{proof}
A formal proof of the equivalence between $S(c)$ and the solution set to the social cost problem \eqref{eq:SWP}, $S^{\text{sc}}(c)$, can be found in \citep{le2020peer}. Here, we show that $S^{\text{sc}}(c)$ has a unique element independent of $c$. First, note that the objective functions $J_i(p_i,c_i)$ are decoupled (namely, depend only on local decision variables) and are strongly convex, for all $c_i$. Hence, \eqref{eq:SWP} has a unique solution $p^*(c)$, that is, $S^{\text{sc}}(c)=\{p^*(c)\}$. Next, we show that $p^*(c)$ is the same for all $c$. Note that, each pair of twin trading terms in $J_i(p_i^*(c_i),c_i)$ and $J_j(p_j^*(c_j),c_j)$, satisfy $c_{(i,j)}^\top p_{ij}^{*,\text{tr}} = - c_{(i,j)}^\top p_{j i}^{*,\text{tr}}$, due to the reciprocity constraints \eqref{eq:RC}. Hence, these terms cancel out when summed up in the objective function \eqref{eq:SWP}, for any choice of $c_{(i,j)}$. It follows that the minimizers of \eqref{eq:SWP} do not depend on $c$.
{\hfill $\square$}
\end{proof}

\begin{rmk}
The dispatch game \eqref{eq:EDPi}, and its optimization counterpart \eqref{eq:SWP}, are the most widespread mathematical formulations for full P2P market designs, and appear with some variations (e.g., trading tariff, price differentiation, and reciprocity) in a number of different work \citep{Baroche:2019, le2020peer, moret2020loss}.
{\hfill $\square$}
\end{rmk}

\subsection{Distributed solution of the P2P trading game} 

\begin{figure}[t] \label{alg:dist_admm}
\hrule
\smallskip
\textsc{Algorithm $1$}: Distributed Peer to Peer Trading
\smallskip
\hrule
\medskip
\textbf{Initialization ($\boldsymbol{k=0}$)}: $p_{ij}^{\text{tr, 0}}, \lambda^{0}_{ij} = 0$, for all trades $(i,j)$.\\[.5em]
\noindent
\textbf{Iterate until convergence:}\\[.3em]
\hspace*{.5em}$
\left\lfloor
\begin{array}{l l}
 & \hspace*{-1em}
\text{For all hubs $i$: }\\[.1em] 
&
\hspace*{-1em}
\left\lfloor
\begin{array}{l}
\text{1. Compute $p^{k+1}_{i}$, $\widehat{p}^{\text{ tr,k+1}}_{ij,i}$ and $\widehat{p}^{\text{ tr,k+1}}_{ij,j}$:}\\ 
\hspace*{1.2em} \text{given $p_{ij}^{\text{tr,k}}$, $\lambda^{k}_{ij}$ , minimize \eqref{eq:lagran} s.t. \eqref{eq:Constr1}-\eqref{eq:RC},} \\[.5em]
\text{2. Broadcast $\widehat{p}^{\text{ tr,k+1}}_{ij,i}$, $\widehat{p}^{\text{ tr,k+1}}_{ji,i}$ to hub $j$, and}
\\ \hspace*{1.2em}
\text{receive $\widehat{p}^{\text{ tr,k+1}}_{ij,j}$, $\widehat{p}^{\text{ tr,k+1}}_{ji,j}$ from $j$, $\forall j\in \mc N\setminus\{i \}$,}\\[.5em]
\text{3. Update trade } p_{ij}^{\text{tr,k+1}} \text{ as in \eqref{eq:averaging}},\\[.5em]
\text{4. Update dual variable } \lambda^{k+1}_{ij} \text{ as in \eqref{eq:dualupdate}},\\
\hspace*{1em}
\vspace*{-1em}
\end{array}
\right.
\\[-.5em]
\hspace*{1em}\\
& \hspace*{-1em} k \leftarrow k+1
\end{array}
\right.
$

\bigskip
\hrule
\end{figure}

To find a variational GNE of \eqref{eq:EDPi}, we use a distributed version of the Alternating Direction Method of Multipliers~(ADMM)~\citep{boyd:2011} on \eqref{eq:SWP}. The resulting algorithm is summarized in Algorithm~$1$. 

The social cost problem in \eqref{eq:SWP} is reformulated as a global consensus problem wherein the hubs have to reach agreement on the bilateral trades. The power traded between hubs, $p^{\text{tr}}_{ij}$,  becomes a global decision variable and each hub $i$ and $j$ optimizes over a local copy of this value, thought of as their local estimate of the trade, $\widehat{p}^{\text{ tr}}_{ij,i}$ and $\widehat{p}^{\text{ tr}}_{ij,j}$, respectively. The economic dispatch problem of each hub $i$ presented in \eqref{eq:EDPi} is therefore solved for the local estimates in addition to $p_i$ subject to additional constraints given below. An iterative consensus procedure is then used to ensure that the local estimates adhere to the global decision,
\begin{subequations}
\label{eq:cons_ADMM}
\begin{align}
  \widehat{p}^{\text{ tr}}_{ij,i}- p_{ij}^{\text{tr}} &= 0,\ \ \  \forall j\in \mc N\setminus\{i \},\\
\widehat{p}^{\text{ tr}}_{ji,i} - p_{ji}^{\text{tr}} &= 0,\ \ \  \forall j\in \mc N\setminus\{i \}. 
\end{align}
\end{subequations}

The augmented Lagrangian for the problem \eqref{eq:EDPi} with the added constraints \eqref{eq:cons_ADMM} is given by

\begin{align}
   \nonumber J_i\left(p_i, c_i\right) &+ \sum_{j\in \mc N\setminus\{i\}} \left( \lambda_{ij}^{\text{T}}(\widehat{p}^{\text{ tr}}_{ij,i}-p_{ij}^{\text{tr}}) + \frac{\rho}{2}\left\|\widehat{p}^{\text{ tr}}_{ij,i}-p_{ij}^{\text{tr}}\right\|_2^2\right.\\ \label{eq:lagran}
    &\left.\lambda_{ji}^{\text{T}}(\widehat{p}^{\text{ tr}}_{ji,i}-p_{ji}^{\text{tr}}) + \frac{\rho}{2}\left\|\widehat{p}^{\text{ tr}}_{ji,i}-p_{ji}^{\text{tr}}\right\|_2^2\right)
\end{align}

where $\lambda_{ij}$ is the Lagrange dual variable and $\rho_{\geq 0}$ is the augmented Lagrangian penalty parameter. The resulting dual function that minimizes \eqref{eq:lagran} subject to the \eqref{eq:Constr1}, and \eqref{eq:RC} is solved independently at the hub level at each iteration to update the local setpoints and estimates of the bilateral trade. Hubs $i$ and $j$ communicate their local estimates of the trade values that are used to update the global trade decision, $p_{ij}^{\text{tr}}$ by an averaging step and the dual variable of \eqref{eq:cons_ADMM} through
\begin{subequations}
\begin{align}
\label{eq:averaging}
\textstyle
  p_{ij}^{\text{tr,k+1}} &= \frac{1}{2} \cdot (\widehat{p}^{\text{ tr,k+1}}_{ij,i} + \widehat{p}^{\text{ tr,k+1}}_{ij,j}),\\
  \label{eq:dualupdate}
\lambda^{k+1}_{ij} &= \lambda^{k}_{ij} + \rho \cdot (\widehat{p}^{\text{ tr,k+1}}_{ij,i} - p_{ij}^{\text{tr,k+1}}).  
\end{align}
\end{subequations}
This process continues until the local and global values of all trades converge, namely, consensus is achieved.

\subsection{Undesired Effects of Autonomous P2P Trading}
In this section, we show that this autonomous peer-to-peer trading model provably leads to a social cost decrease, but not necessarily to a local cost decrease for each hub.

The economic dispatch problem without bilateral trading corresponds to \eqref{eq:EDPi} with $p^{\text{tr}}_{ij}=0$ for all hubs. Clearly, the corresponding social cost, $W^{\text{nt}}(c)$, will be greater than $W(c)$ in \eqref{eq:SWP}, since the feasible set of the non-trading scenario $\mathscr{P}^{\text{nt}}=\mathscr{P}\cap\{p~|~ p^{\text{tr}}_{ij}=0, \; \forall j\in \mc N\setminus\{i\}, \, \forall i \in \mc N\}$ is a subset of the feasible set with bilateral trades $\mathscr{P}$. Hence, allowing bilateral trades cannot increase the social cost, regardless of the trading prices.

The decrease in social cost is not necessarily reflected in the individual costs of all agents. In other words, there may exist profiles of bilateral trading prices for which certain hubs are worse off than if they were not participating in the autonomous trading mechanism. We illustrate this phenomenon via a numerical example for a 3-hub network in Section \ref{sec:simulation}.
%
%
A natural question is whether there exists a price profile for which all the hubs benefit by participating in the autonomous bilateral trading mechanism. The following theorem gives an affirmative answer to this question.
\begin{thm}
\label{lmm:feasibility}
Let $ p^*$ be the unique variational GNE of \eqref{eq:EDPi}. There exists a price profile $ c^* \in \R^{N(N-1)H}$ such that 
\begin{align}
\label{eq:SGC}
J_i( p_i^*,c_i^*) \leq  J_i^{\text{nt}}, \quad \forall i \in \mc N,
\end{align}
where $J_i^{\text{nt}}$ are the local costs of the non-trading scenario.
\end{thm}
\begin{proof}
Without loss of generality, we carry out the proof for $H=1$. Consider the unique variational GNE $ p^*$ and label all $E$ realized trades between hubs in $ p^*$ as $t_\ell$, for $\ell \in\mc E =\{1,\dots,E\}$. Then, define a matrix $V\in \bb R^{N \times E}$, whose $(\ell, m)$-entry satisfies, for all $m \in \mc N$, $\ell \in \mc E$:
\begin{equation*} \textstyle
[V]_{  m \ell} :=
\begin{cases} \textstyle
\bar p^{\textrm{tr},*}_{ij}  & \text{if } t_\ell=(i,j) \text{ and } m=i, \\ \textstyle
\bar p^{\textrm{tr},*}_{ji} & \text{if } t_\ell=(i,j) \text{ and } m=j, \\ \textstyle
0 & \text{otherwise},
\end{cases}
\end{equation*}
Since $ p^{\textrm{tr},*}_{ij}= - p^{\textrm{tr},*}_{ji}$ by the trading reciprocity \eqref{eq:RC}, $V$ is indeed a graph incidence matrix and satisfies
\begin{align}
\label{eq:conncected}
\textrm{range}(V) \supseteq \textrm{range}(VV^\top) = \textrm{null}\mathbf(1_N^\top),
\end{align}
where the equality follows from \citep[Th. 8.3.1]{godsil2013algebraic} assuming $V$ describes a connected (trading) network. If the network is not connected, the remainder of the proof can be carried out for each connected component.\\
Since the social cost of $p^*$, $W$, is no greater than that of the non-trading case $W^{\text{nt}}$, we can write
\begin{align}
\label{eq:sproof}
W - W^{\text{nt}} + \kappa = 0, \text{ for some }  \kappa\geq 0,
\end{align}
Now, define $\bb J = \col(J_1,\ldots,J_N)$, $\bb J^{\text{nt}} = \col(J_1^{\text{nt}},\ldots,J_N^{\text{nt}})$. Then, it holds that
$
\mathbf 1_N^\top\big(\bb J - \bb J^{\text{nt}} + \frac{\kappa}{N}\mathbf{1}_N \big) = 0, 
$
which implies 
\begin{align}
\label{eq:sproof2}
\textstyle
\bb J  - \bb J^{\text{nt}} + \frac{\kappa}{N}\mathbf{1}_N \in \textrm{null}(\mathbf 1_N^\top) \subseteq \textrm{range}(V),
\end{align}
where for the last inclusion we used \eqref{eq:conncected}. It follows by \eqref{eq:sproof2} that there exists a vector $c^*$ such that
\begin{align}
\label{eq:sproof3}
\textstyle
\bb J  + V c^*  =  \bb J^{\text{nt}} - \frac{\kappa}{N}\mathbf{1}_N \leq \bb J^{\text{nt}},
\end{align}
where the $i$-th component of \eqref{eq:sproof3} is in fact \eqref{eq:SGC}.
{\hfill $\blacksquare$}
\end{proof}

In the following section, we develop a scalable mechanism to identify bilateral trading prices that are not only locally beneficial for each hub but also fair. 

\section{Designing Fairness via Bilevel Games}
To ensure that the equilibrium of the autonomous peer-to-peer trading game \eqref{eq:EDPi} is ``fair", we set the prices of the bilateral trades by solving the following bilevel game
\begin{subequations}%
\label{eq:bilevel}
\begin{align}
\label{eq:UL}
\min_{c , \,p} \quad  & \varphi\left(p,c\right) \\
\text { s.t. } ~~ & 
c \in \mc C,\\
\quad 
&p \in S(c), \label{eq:LL1}
\end{align}
\end{subequations}
where $\mc C $ is a set of feasible trading prices that can be used to model price regulations, such as capping, and $S$ as in Lemma~\ref{lmm:Sens}.
The lower level \eqref{eq:LL1} imposes the operational setpoints and the bilateral trades $p$ to be in the variational GNE set $S(c)$ of the price-parametrized game \eqref{eq:EDPi}. At the higher level, the optimal trading prices are chosen to minimize a certain \textit{fairness metric} $\varphi$ which depends also on the operational setpoints of the lower level.

Here, we define the fairness metric as the sample variance of the normalized cost reductions $d_i$ achieved by enabling peer-to-peer trading between hubs, namely
\begin{align}
\label{eq:FM}
\varphi(p,c)= \sum_{i\in \mc N}\left(d_i(c_i,p_i)-
\frac{1}{N}\sum_{j\in \mc N} d_j(c_j,p_j)\right)^2,
\end{align}
where the normalized cost reduction $d_i$ is defined as
\begin{align}
\label{eq:cost_red}
d_i\left(c_i,p_i\right)=(J_i^{\text{nt}}-J_i\left(p_i, c_i\right))/J_i^{\text{nt}}, \quad \forall i \in \mc N.
\end{align}
The non-trading cost $ J_i^{\text{nt}}$ can be locally computed by each hub by solving their optimal economic dispatch problem in which trading is disabled. This metric ensures that the social wealth generated by enabling peer-to-peer trading is ``fairly" distributed across the hubs.

Large-scale bilevel games as \eqref{eq:bilevel} are notoriously difficult to solve. Existing solution approaches are based on mixed-integer programming or nonlinear relaxations, and lack either convergence guarantees or computational efficiency. Next, we show that a solution to \eqref{eq:bilevel} can instead be efficiently obtained by solving sequentially the game \eqref{eq:LL1} and the optimization \eqref{eq:UL}.
By Lemma~\ref{lmm:Sens}, the v-GNEs set $S(c)$ is a singleton, independent on the bilateral trading prices $c$. Hence, the bilevel game \eqref{eq:bilevel} boils down to 
\begin{align}
\label{eq:eqSL2}
\min_{c } \quad  & \varphi\left(p^*,c\right) \quad  \text{s.t.} \quad   c \in \mathcal C,
\end{align}
where $p^*$ is the unique price-independent v-GNE of \eqref{eq:EDPi}.
Hence, a solution to \eqref{eq:bilevel} can be found in two steps:
\begin{enumerate}[1.]
\item Compute the unique v-GNE $p^*$ of  \eqref{eq:EDPi}, for any fixed $c$;

\item Compute a solution to \eqref{eq:eqSL2},
where $p^*$ is fixed.
\end{enumerate}

Solving \eqref{eq:eqSL2} centrally requires global knowledge over the optimal operational setpoints $p^*=(p^*_1,\ldots,p^*_N)$, the cost functions $J_i$, and the non-trading cost, $J_i^{\text{nt}}$, which is unrealistic. Additionally, the dimensionality of \eqref{eq:eqSL2} increases quadratically with the number of hubs, thus making it rapidly intractable for large-scale hub networks. Motivated by these challenges, in the next section, we design a scalable and privacy-preserving algorithms to solve \eqref{eq:eqSL2}.

\begin{rmk} The single-level reformulation \eqref{eq:eqSL2} is obtained by exploiting the specific structure of the market model \eqref{eq:EDPi}, for which the variational Nash equilibrium is price insensitive (by Lemma~\ref{lmm:Sens}). Variations of this market design that consider price differentiation \citep{Sorin:2019} or trading reciprocity constraints with inequalit y\citep{le2020peer} do not enjoy this favourable property.
{\hfill $\square$}
\end{rmk}

\subsection{A Semi-decentralized Price Mediation Protocol}
We design a semi-decentralized projected-gradient algorithm which preserves privacy and achieve scalability with respect to the number of hubs.
%
To distribute the computation, a mediator, $M_{(i,j)}$, is introduced between each pair of hubs $i$ and $j$, whose objective is to determine a fair trading price, $c_{(i,j)}$. The mediator updates the price according to
\begin{equation}
\label{eq:grad_descent}
c_{(i,j)}^{k+1}=\Pi_{\mathcal{C}}\left( c_{(i,j)}^k-\beta \frac{\partial \varphi(p^*,c^k)}{\partial c_{(i,j)}}\right),
\end{equation}
where $\Pi_{\mathcal{C}}$ is the projection onto a feasible set of prices $\mathcal{C}$. The gradient of $\varphi(p,c)$ with respect to a price $c_{(i,j)}$ is 
\begin{align*}
 \frac{\partial \varphi(p,c)}{\partial c_{(i,j)}}
 =&\textstyle 
 \frac{2}{N}
  \left( \frac{p^{\text{tr}}_{ij}}{J_i^{\text{nt}}}\left(d_i(c_i,p_i)\!-\!\bar d\left(c,p\right)\right) \right.
  \\ & \quad \textstyle
   \left. + \frac{p^{\text{tr}}_{ji}}{J_j^{nt}}\left(d_j\left(c_j,p_j\right)\!-\!\bar d\left(c,p\right)\right)\right)   
\end{align*}
where $\bar d(c,p) = \frac{1}{N}\sum_{i=1}^N d_i\left(c_i,p_i\right)$ is the average cost reduction.
\begin{figure}
\begin{center}
\includegraphics[width=7.5cm]{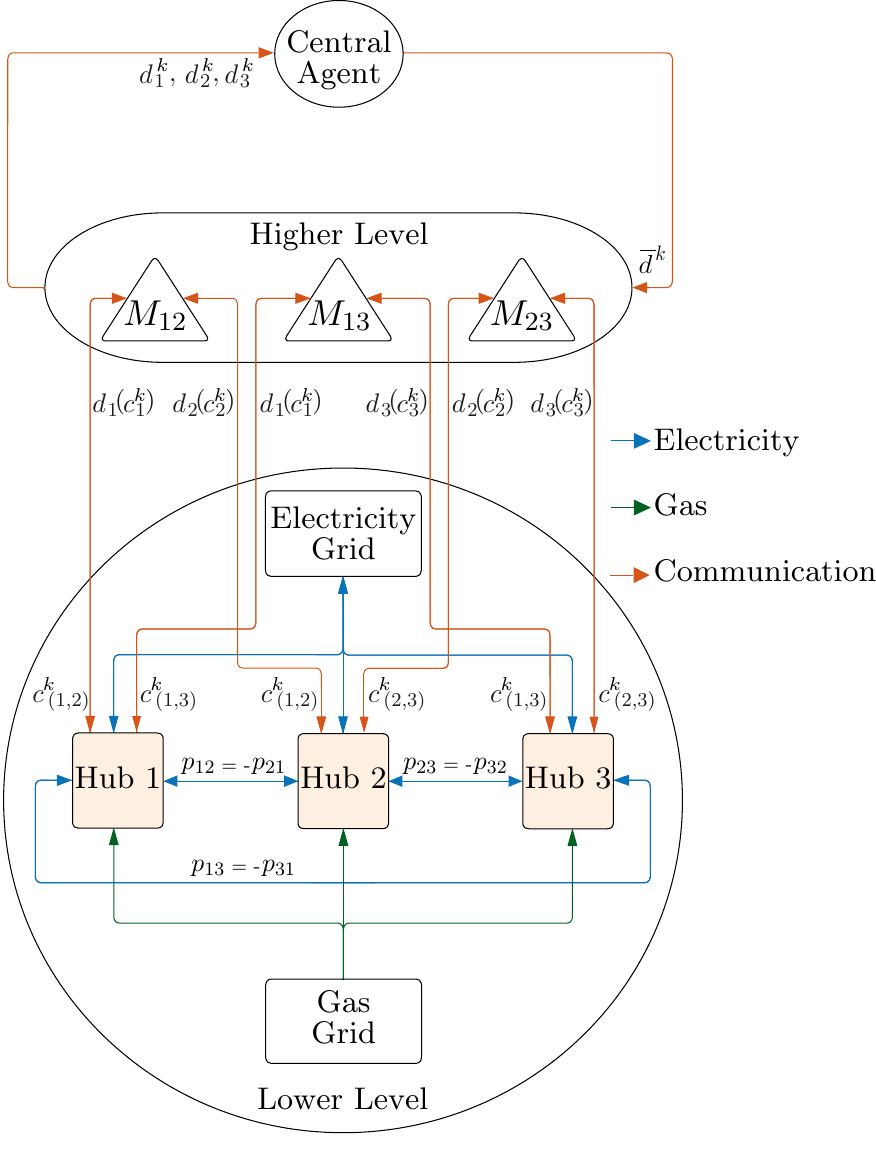}
\caption{Flowchart of the price mediation scheme in Alg.~2.} 
\label{fig:bilevel_alg}
\end{center}
\end{figure}
A central coordinator is introduced that gathers the $\bar d_i$ and broadcasts $\bar d$ to all the mediators. Algorithms with this information structure are called semi-decentralized, see e.g. \citep{belgioioso2021semi}.

The resulting algorithm is summarized in Algorithm~2 and its information structure is illustrated in Figure~\ref{fig:bilevel_alg}. 
At every iteration, each mediator $M_{ij}$ receives the normalized cost reductions, $d_{i}$ and $d_{j}$, from the hubs it manages, and the average network cost reduction $\bar d$ from the coordinator. Then, it updates the price $c_{(i,j)}$ with a projected-gradient step \eqref{eq:grad_descent}.
This process continues until the prices of all trades reach convergence. Since the objective function  $\varphi \left(p,c\right)$ is convex and $L$-smooth\footnote{The convexity of $\varphi(\cdot, p)$ can be proven by showing that its Hessian is positive semidefinite. A formal proof is omitted here due to space limitation; smoothness follows since $\phi(\cdot,p)$ is quadratic.}, for some $L>0$, uniformly in $p$, taking the step size $\beta \in (0, 2/L)$, in the price update \eqref{eq:grad_descent}, guarantees convergence of Algorithm 2. 
\begin{figure}[t] \label{algo:PriceMediation}
\hrule
\smallskip
\textsc{Algorithm $2$}: Price Mediation Mechanism
\smallskip
\hrule
\medskip
\textbf{Initialization ($\boldsymbol{k=0}$)}: $c^0_{(i,j)} = 0$, for all trades $(i,j)$.

\noindent

\medskip
\textbf{Iterate until convergence:}\\[.5em]
\hspace*{.5em}$
\left\lfloor
\begin{array}{l l}
 & \hspace*{-1em}
\text{For all  mediators $M_{ij}$: }\\[.5em] 
&
\hspace*{-1em}
\left\lfloor
\begin{array}{l}
\text{1. Receives $d_\ell(c_\ell^k)$  from each hub $\ell \in \{i,j \}$,} \\[.5em]
\text{2. Receives  $\bar d^k$ from the coordinator,} \\[.5em]
\text{3. Update price } c^{k+1}_{(i,j)} \text{ as in \eqref{eq:grad_descent}},
\end{array}
\right.
\\[1em]
\hspace*{1em}\\
 &
 \hspace*{-1em} \text{Coordinator:  }\\[.5em]
&
\hspace*{-1em} \textstyle
\left\lfloor
\begin{array}{l}
\text{1. Gather } \bar d^{k+1} = \sum_{i \in \mc N} d_i^k\left(c_k\right),
\\[0.5em]
\text{2. Broadcast $\bar d^{k+1}$ to all mediators,}  \\[-.5em]
\hspace*{1em}
\end{array}
\right.
\\[-.5em]
\hspace*{1em}\\
& \hspace*{-1em} k \leftarrow k+1
\end{array}
\right.
$

\bigskip
\hrule
\end{figure}
%
%
%
%
%
%
%


\section{Numerical Results}
\label{sec:simulation}
We illustrate the proposed price mediation mechanism on a 3-hub network. The configuration, parameters and capacities for the three hubs are presented in Appendix~\ref{Appendix:A}. In general, the cost functions for electricity and gas are linear and the v-GNE is not unique. In this study, an additional regularization term that minimizes the total energy imported is added to the cost to find a unique solution. Alternatively, selection algorithms can be used instead of regularization to handle the non-uniqueness of the v-GNE~\citep{Ananduta:2022}. The price of input energy carriers (electricity and gas) and for utilizing the electricity grid are summarized in Table~\ref{tab:tariff}. We solve the optimization for a horizon $H = \SI{24}{\hour}$ with a sampling resolution of $\SI{1}{\hour}$.

 \begin{table}[H]
        \centering
    \centering
    \begin{tabular}{l c c}
                \hline
                Tariff & Price(CHF/kW) \\ \hline
                Electricity output  & 0.22\\
                Electricity feed-in  & 0.12\\ 
                Gas  & 0.115\\ [1em]
        \end{tabular}
        \caption{Tariffs for electricity and gas utility.}
                \label{tab:tariff}  
\end{table}

\begin{figure}
\begin{center}
\includegraphics[width=8.4cm]{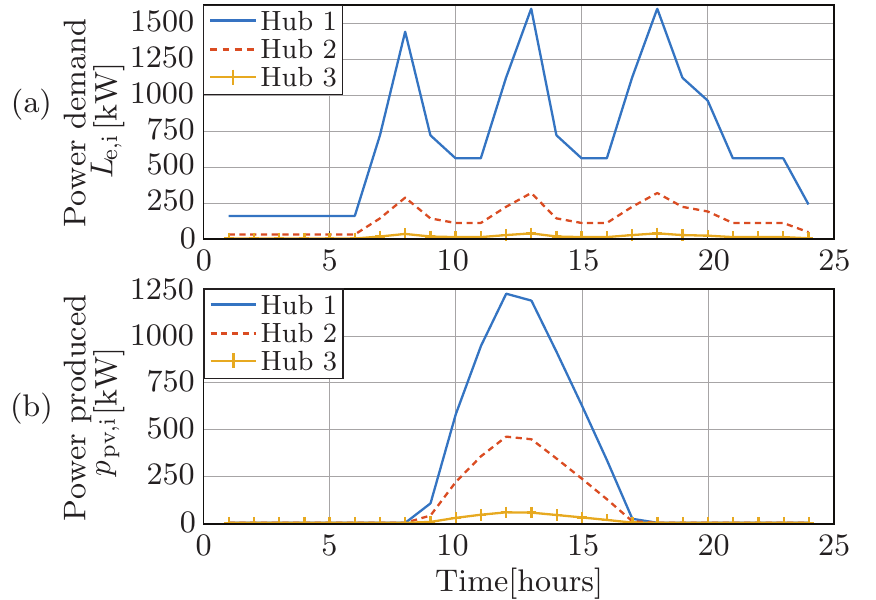}
\caption{Electric power demand, $L_{\text{e,i}}$, and the PV production, $p_{\text{pv,i}}$, for the three hubs($i = 1,2,3$) over 24 hours.} 
\label{fig:input}
\end{center}
\end{figure}

The electricity demand and PV production for the three hubs over a span of ${24}$ hours are shown in Figures \ref{fig:input}(a) and (b), respectively. Hub 1 represents a larger industrial hub with a high production capacity, Hub 2 is a medium sized hub, and Hub 3 is a small residential hub with heat pump, PV, and small energy demand. 
\subsection{The impact of autonomous peer-to-peer trading}

First, we compare the performance of the system without and with autonomous peer-to-peer energy trading.
All the bilateral trades and the operational setpoints of the converters are calculated by solving \eqref{eq:EDPi} with Algorithm~1. The results are illustrated in Figure~\ref{fig:powertrading}(a). 

At the beginning of the day (0:00–7:00) and at end of the day (18:00-24:00), when there is no PV production, power is traded from the larger Hub~1 that can use CHP to produce electricity to Hub~2 and Hub~3, owing to its larger production capacity. In the absence of trading, Hub~2 and Hub~3 import electricity from the grid at a higher price than the price of the gas used to produce the energy in the CHP. Part of the power traded from Hub~1 to Hub~3 is transferred to Hub~2 to minimize the grid tariff levied to the hubs; this grows quadratically for the power transferred between any two hubs, making it profitable to make multiple small power trades than a single large one. When PV output is high, the trades drop to 0 and the cost is equivalent to that in the non-trading case since the PV output of each hub is sufficient to fulfil the local demand. 

\subsection{The impact of the bilateral trading prices}
We verify that the optimal power traded between the hubs and the optimal setpoints, $p^*$, are independent of the trading price (Lemma~\ref{lmm:Sens}) by solving \eqref{eq:EDPi} with different trading prices.
Figure~\ref{fig:cost_nonoptimal_optimal} shows the cost reduction \eqref{eq:cost_red} achieved by each hub for three different trading prices (uniform across all peer-to-peer trades), namely, $c = 0.1, 0.18, 0.2$ CHF/kWh. The figure also shows the reduction in the social cost compared to when no trading occurs, and the benefit of autonomous trading to the hub network is evident by the 2.5\% reduction of social cost, independently of the trading price.
For $c = 0.1$ CHF/kWh, since the trading price is low (even lower than the feed-in tariff), Hub~2 and Hub~3 benefit by trading as they import cheap energy from Hub~1. However, this results in an increase of the cost for Hub~1, as the trading price does not cover the additional production costs of the power traded to the other hubs. For $c = 0.18$ CHF/kWh, although each of the hubs benefits from trading, the cost reduction varies drastically between the hubs. Hub~1 that exports much of its energy has a much smaller benefit than Hub~2 that only imports energy. Finally, for $c = 0.2$ CHF/kWh, Hub~1 and Hub~2 continue to benefit whereas the smaller Hub~3 loses since the higher trading price for import and the trading tariff increase the net price to more than the grid price. 

\begin{figure}
\begin{center}
\includegraphics[width=8.4cm]{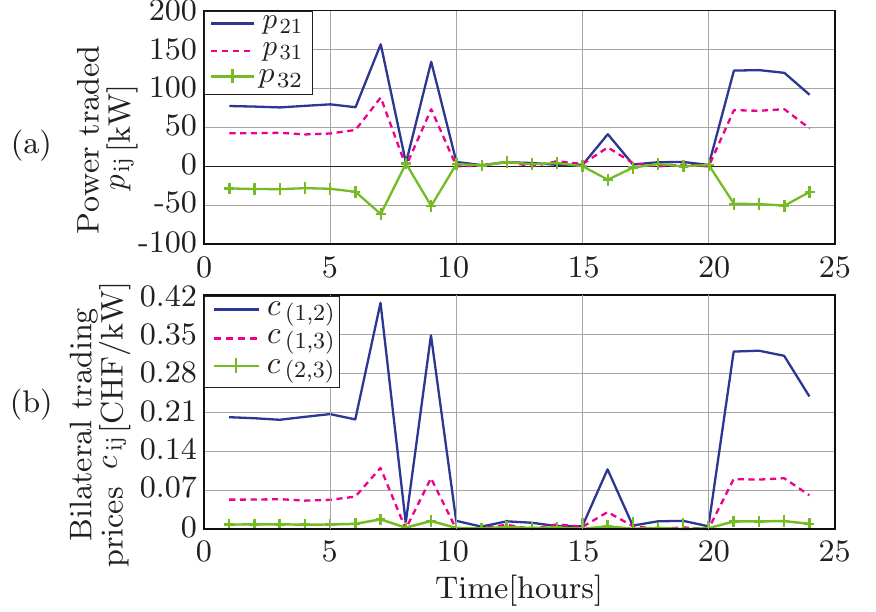}
\caption{(a) Bilateral power traded  and (b) trading price for the bilateral trades in the 3-hub network.} 
\label{fig:powertrading}
\end{center}
\end{figure}

\begin{figure}
\begin{center}
\includegraphics[width=8.4cm]{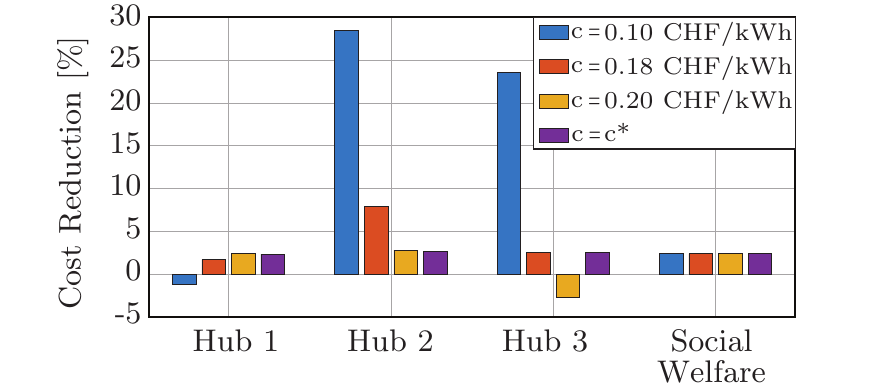}
\caption{Cost reduction \eqref{eq:cost_red} for three different trading prices and the optimized prices $c^*$ illustrated in Figure~\ref{fig:powertrading}(b).
} 
\label{fig:cost_nonoptimal_optimal}
\end{center}
\end{figure}

%
Finally, the trading prices are calculated using the price mediation mechanism in Algorithm~2. The resulting cost reduction for the three hubs using the optimal price profile is shown in Figure~\ref{fig:cost_nonoptimal_optimal}. Interestingly, the trading price found by Algorithm~2 is different for each trade and also varies at different times within the control horizon (shown in Figure~\ref{fig:powertrading}(b)). The normalized cost reduction achieved by each of the hubs is nearly equal and matches the social cost reduction achieved by the network.

\section{Conclusion}
Energy trading prices play a major role in incentivizing participation in autonomous peer-to-peer trading mechanisms. We proposed a privacy-preserving and scalable price-mediation algorithm that provably finds price profiles that are not only locally-beneficial for each hub but also network-wide fair. Numerical simulation on a 3-hub network supported this theoretical result.

\bibliography{biblio}
\clearpage
\appendix
\section{Parameters for numerical simulations} 
\label{Appendix:A}
   \begin{table}[b]
        \centering
    \centering
    \begin{tabular}{l c c}
                \hline 
                \multicolumn{3} {c} {\textbf{Hub 1}}\\ \hline
                \multicolumn{2}{l}{Component\qquad Parameter} & Value \\ \hline
                &$\eta_{\text{chp,1}}$ &0.36\\
                CHP & [$p_\text{A,1}$, $p_\text{B,1}$,$p_\text{C,1}$,$p_\text{D,1}$] &
              [380, 315, 745, 800]~kW 
                \\
                 & [$q_\text{A,1}$, $q_\text{B,1}$,$q_\text{C,1}$,$q_\text{D,1}$]&[0, 515, 1220, 0]~kW\\ 
                HP & COP, [$q_{\text{hp},1}^{\text{min}}$, $q_{hp,1}^{\text{max}}$] & 4.5, [0, 450]~kW\\ 
                GB  & $\eta_{\text{gb},1}$, [$q_{\text{gb},1}^{\text{min}}$, $q_{\text{gb},1}^{\text{max}}$] & 0.78, [0, 350]~kW\\
                PV &$\eta_{\text{pv},1}$, $a_{\text{pv},1}$, [$p_{\text{pv},1}^{\text{min}}$, $p_{\text{pv},1}^{\text{max}}$]&0.15, $\SI{8400}{\meter}^2$, [0, 2500]~kW\\ 
                ES  & $\eta_{\text{e},1}$, $\gamma_{\text{e},1}$, [$p_{\text{s},1}^{\text{min}}$, $p_{\text{s},1}^{\text{max}}$]&0.99, 0.999, [50, 750]~kWh\\
                & [$p_{\text{ch},1}^{\text{min}}$,$p_{\text{ch},1}^{\text{max}}$], [$p_{\text{dc},1}^{\text{min}}$, $p_{\text{dc},1}^{\text{max}}$] &[0,200]~kW, [0,200]~kW\\
                TS  & $\eta_{\text{h},1}$,$\gamma_{\text{h},1}$, $q_{\text{s},1}^{\text{min}}$, $q_{\text{s},1}^{\text{max}}$&0.95,0.992,[290, 12900]~kWh\\
                & [$q_{\text{ch},1}^{\text{min}}$,$q_{\text{ch},1}^{\text{max}}$], [$q_{\text{dc},1}^{\text{min}}$, $q_{\text{dc},1}^{\text{max}}$] &[0,3200]~kW, [0,3200]~kW \\[1em] \hline 
                \multicolumn{3} {c} {\textbf{Hub 2}}\\ \hline
                \multicolumn{2}{l}{Component\qquad Parameter} & Value \\ \hline
                PV  &$\eta_{pv,2}$, $a_{pv,2}$,[$p_{\text{pv},2}^{\text{min}}$, $p_{\text{pv},2}^{\text{max}}$]&0.15, $\SI{3170}{\meter}^2$, [0,350]~kW\\ 
                HP & COP, [$q_{\text{hp},2}^{\text{min}}$, $q_{hp,2}^{\text{max}}$] & 4.5, [0,300]~kW\\
                GB  & $\eta_{gb,2}$, [$q_{gb,2}^{\text{min}}$, $q_{gb,2}^{\text{max}}$] & 0.78, [0,50]~kW\\
                TS  & $\eta_{\text{h},2}$,$\gamma_{\text{h},2}$, [$q_{\text{s},2}^{\text{min}}$, $q_{\text{s},2}^{\text{max}}$]&0.95,0.992,[0.36, 1.62]~kWh\\
                & [$q_{\text{ch},2}^{\text{min}}$,$q_{\text{ch},2}^{\text{max}}$], [$q_{\text{dc},2}^{\text{min}}$, $q_{\text{dc},2}^{\text{max}}$] &[0,0.3]~kW, [0,0.3]~kW \\[1em]
                 \hline
                \multicolumn{3} {c} {\textbf{Hub 3}}\\ \hline
                \multicolumn{2}{l}{Component\qquad Parameter} & Value \\ \hline
                PV  &$\eta_{pv,3}$, $a_{pv,3}$, [$p_{\text{pv},3}^{\text{min}}$, $p_{\text{pv},3}^{\text{max}}$]&0.15, $\SI{380}{\meter}^2$, [0, 80]~kW\\ 
                HP & COP, [$q_{\text{hp},3}^{\text{min}}$, $q_{hp,3}^{\text{max}}$] & 4.5, [0, 50]~kW\\ [1em]
        \end{tabular}
        \caption{Parameters and capacities for energy hubs used in the numerical study.}
                \label{tab:ehub_parameters}  
\end{table}
\end{document}